\documentstyle{article}

\font\tenrm=cmr10
\font\tenit=cmti10
\font\elevenbf=cmbx10 scaled\magstep 1
\font\elevenrm=cmr10 scaled\magstep 1
\font\elevenit=cmti10 scaled\magstep 1

\catcode`\@=11 % This allows us to modify PLAIN macros.
\def\alt{\mathrel{\mathpalette\@versim<}}
\def\agt{\mathrel{\mathpalette\@versim>}}
\def\@versim#1#2{\vcenter{\offinterlineskip
        \ialign{$\m@th#1\hfil##\hfil$\crcr#2\crcr\sim\crcr } }}

\renewenvironment{thebibliography}[1]
 { \elevenrm
   \begin{list}{\arabic{enumi}.}
    {\usecounter{enumi} \setlength{\parsep}{0pt}
     \setlength{\itemsep}{3pt} \settowidth{\labelwidth}{#1.}
     \sloppy
    }}{\end{list}}

\begin{document}
\begin{center}
\vglue 0.6cm
{ {\elevenbf        \vglue 10pt
               RENORMALIZATION GROUP STUDY\\
               \vglue 3pt
         OF THE MINIMAL SUPERSYMMETRIC STANDARD MODEL:\\
               \vglue 3pt
         NO SCALE MODELS
%%%\footnote{
%%%\ninerm\baselineskip=11pt This text was produced using \LaTeX\.}
\\}
\vglue 5pt
%%%{\ninerm (For 20\% Reduction to 8.5 $\times$ 6 in Trim Size)\\}
\vglue 1.0cm
{\tenrm PIERRE RAMOND\\}
\baselineskip=13pt
{\tenit Institute for Fundamental Theory and
Physics Department,\\}
\baselineskip=12pt
{\tenit University of Florida, 215 Williamson Hall\\}
\baselineskip=12pt
{\tenit Gainesville, Florida 32611, U.S.A.\\}}

\vglue 0.8cm

{\tenrm ABSTRACT}
\end{center}
\vglue 0.3cm
{\rightskip=3pc
\leftskip=3pc
\tenrm\baselineskip=12pt
\noindent

We present partial numerical results in the Minimal Supersymmetric Standard
Model with soft breaking of supersymmetry, and radiative breaking of the
electroweak symmetry. We impose the additional relation $m_b=m_\tau$ at the GUT
scale. For the special case of the strict no-scale model, in which global
supersymmetry breaking arises solely from soft gaugino masses, we find that
$M_t$ is expected to be lighter than $\sim 128$ GeV.  A higher upper bound for
$M_t$ of $\sim 138$ GeV is predicted, if $M_b$ is at the lower end of its
experimental uncertainty.  }

%Typeset the abstract with an indentation
%of 3 picas on the left and right margins and in 10 point roman with
%baselineskip of 12 point.}

\vglue 0.6cm
{\elevenbf\noindent 1. Introduction}
\vglue 0.4cm
\baselineskip=14pt
\elevenrm

In this paper, I report some preliminary results of work in progress, done in
collaboration with E. Piard and D. Casta\~ no. In the last few years, it has
become apparent, using the ever increasing accuracy in the measurement of the
strong coupling, that Supersymmetry (SUSY) affords an elegant means to achieve
gauge coupling unification \cite{amaldi,ekn,langluo} at scales consistent with
Grand Unified Theories (GUTs) \cite{patisalam,geoglash,georgi,frimink,gurasi}.
Whereas in the Standard Model (SM) the three gauge couplings unify ``two by
two'' forming the ``GUT triangle,'' in the simplest Minimal SUSY Extension of
the SM (MSSM) these gauge couplings spectacularly unify at a point (within the
experimental errors in their values).  Given that the scale of unification in
these models is generally above the lower bound set by proton decay, these
supersymmetric grand unified theories (SUSY-GUTs) have regained increasing
interest.  Constraints from the Yukawa sectors of such models have also yielded
interesting predictions for various low energy parameters including the top
quark mass \cite{summer91,kln,anan,giveon}.

Crucial to these analyses has been the use of the Renormalization
Group (RG) in extrapolating from the GUT scale to experimental scales.
Here, we use of our previous RG analysis for the standard
model\cite{big},
including the supersymmetric two loop $\beta$ functions for the MSSM.
Also soft symmetry breaking (SSB) terms are added. These lead
to a nondegenerate superparticle spectrum as well as to the radiative
breaking of the electroweak symmetry.  A brief discussion of the
effective one loop potential is presented.
The boundary
conditions at the unification scale in these minimal low energy supergravity
models are discussed.  Next the numerical procedure employed is described.
The treatment of thresholds and the ``special'' form of the $\beta$
functions needed is discussed next.  Similar analyses have appeared in the
literature \cite{rossrob,klpny,an,na}.

In the following, we only concern ourselves with the simplest schemes which
have a chance of reproducing the data. If these are indeed those chosen by
nature, they have some very distinct signatures. The top
quark, although allowed by the standard model to be as heavy as $200$ GeV,
comes out much lighter. The lightest Higgs particle also comes out
light, so light in fact that it may be very hard to detect. Another feature is
that these models all predict the existence of a stable particle, the LSP,
whose
abundance can be calculated, based on the standard cosmological scenario. LSPs
are typically too light to bring $\Omega$ to one, the
preferred value of theorists.
\vglue 0.5cm
{\elevenbf\noindent 2. Minimal Supersymmetric Standard Model }
\vglue 0.4cm

In the minimal supersymmetric extension of the standard model, every particle
has a supersymmetric partner, with spin differing by a half \cite{nilles}.
Supersymmetry requires a second Higgs field with opposite hypercharge to the
first as the superpotential cannot contain both a field and its complex
conjugate.  The second Higgs is also needed for both chiral $U(1)$ and $SU(2)$
global anomaly cancellation and to give its sector a mass.  For renormalizable
theories, the superpotential can have at most degree three interactions.  The
superpotential for the MSSM is (suppressing the $SU(2)$ and Weyl metrics)
\begin{equation}
   W = {\hat{\overline u}} {\bf Y}_u {\hat \Phi}_u {\hat Q} +
       {\hat{\overline d}} {\bf Y}_d {\hat \Phi}_d {\hat Q} +
       {\hat{\overline e}} {\bf Y}_e {\hat \Phi}_d {\hat L} +
          \mu {\hat \Phi}_u {\hat \Phi}_d \ .
\label{superpotential}
\end{equation}
where the hat indicates a chiral superfield and the overline denotes a left
handed $CP$ conjugate of a right handed field,
${\overline\psi}=i\sigma_2\psi_R^*$.  The usual Yukawa interactions are
accompanied by new Yukawa interactions among the scalar quarks and leptons and
the Higgsinos in the supersymmetric Lagrangian.  There are also new gauge
Yukawa interactions involving the gauginos.  The new purely scalar interactions
form the scalar potential which is positive definite in supersymmetric
theories.  The scalar potential will be discussed in a subsequent section. A
remarkable aspect of supersymmetry is that all these new interactions require
no new couplings. Table I displays the $SU(3)\times SU(2)\times U(1)$ quantum
numbers of the chiral (all left handed) and vector superfields of the MSSM.

The last term is put in by hand to avoid the exact Peccei-Quinn (PQ) symmetry
that the superpotential would otherwise exhibit. Although there are other ways
to treat the PQ symmetry, including producing an axion, we leave the other
possibilities to a future investigation.
%--------------------------------------------------------------------------
\begin{table}
\caption{Quantum Numbers}
\begin{center}
\begin{tabular}{rrrrrrrrrrrrr}
\hline
  & ${\hat Q}$ & ${\hat{\overline u}}$ & ${\hat{\overline d}}$ &
${\hat L}$ & ${\hat{\overline e}}$ & ${\hat\Phi_u}$ & ${\hat\Phi_d}$ &
${\hat g}^A$ & ${\hat W}^a$ & ${\hat B}$ \\
\hline
$U(1)$ & $+{1\over6}$ & $-{2\over3}$ & $+{1\over3}$ & $-{1\over2}$ &
$+1$   & $+{1\over2}$ & $-{1\over2}$ & $0$ & $0$ & $0$ \\
$SU(2)$ & ${\bf 2}$ & ${\bf 1}$ & ${\bf 1}$ & ${\bf 2}$ & ${\bf 1}$ &
${\bf 2}$ & ${\bf 2}$ & ${\bf 1}$ & ${\bf 3}$ & ${\bf 1}$ \\
$SU(3)$ & ${\bf 3}$ & ${\overline{\bf 3}}$ & ${\overline{\bf 3}}$ &
${\bf 1}$ & ${\bf 1}$ & ${\bf 1}$ & ${\bf 1}$ & ${\bf 8}$ &
${\bf 1}$ & ${\bf 1}$ \\
\hline
\end{tabular}
\end{center}
\end{table}
%--------------------------------------------------------------------------

\vglue 0.5cm
{\elevenbf \noindent 3. Minimal Low Energy Supergravity Model}
\vglue 0.4cm

Since no super particles have been observed experimentally, supersymmetry,
if truly present in nature, must be broken.  One way to accomplish this
breaking is to add to the Lagrangian soft breaking terms.

The most general soft symmetry breaking potential for the MSSM can be written
(including gaugino mass terms)
%\begin{mathletters}
\begin{eqnarray}
   V_{soft} &=&  m_{\Phi_u}^2 \Phi_u^\dagger \Phi_u^{}
               + m_{\Phi_d}^2 \Phi_d^\dagger \Phi_d^{}
               + B \mu ( \Phi_u \Phi_d + h.c. ) \nonumber \\
   & &\mbox{}  + \sum_i\bigl(\;
                 m_{Q_i}^2 {\tilde Q}^\dagger {\tilde Q}
               + m_{L_i}^2 {\tilde L}^\dagger {\tilde L}
               + m_{u_i}^2 {\tilde{\overline u}}^\dagger
                           {\tilde{\overline u}}
               + m_{d_i}^2 {\tilde{\overline d}}^\dagger
                           {\tilde{\overline d}}
               + m_{e_i}^2 {\tilde{\overline e}}^\dagger
                           {\tilde{\overline e}} \;\bigr) \nonumber \\
    & &\mbox{} + \sum_{i,j}\bigl(\;
                 A_u^{ij}Y_u^{ij}{\tilde{\overline u}}_i\Phi_u{\tilde Q}_j
               + A_d^{ij}Y_d^{ij}{\tilde{\overline d}}_i\Phi_d{\tilde Q}_j
               + A_e^{ij}Y_e^{ij}{\tilde{\overline e}}_i\Phi_d{\tilde L}_j
               + h.c.
                 \;\bigr) \ , \label{vsoft} \\
    V_{gaugino} &=& {1\over2} \sum_{l=1}^3
                M_l \lambda_l \lambda_l + h.c. \ , \label{vgaugino}
\end{eqnarray}
%\end{mathletters}
\noindent
where $V_{gaugino}$ is the Majorana mass terms for the gaugino fields,
$\lambda_l$ (suppressing the group index), corresponding to $U(1)$, $SU(2)$,
and $SU(3)$, respectively. In the above, there are in principle sixty-three
soft symmetry breaking parameters. This is hardly desirable to explain the
standard model which already has eighteen parameters! It is clear that we need
a further organizing principle. While this may be done by invoking extra
symmetries such as flavor blindness (blandness), one attractive possibility is
to couple the standard model to $N=1$ supergravity (SUGRA).  In the minimal low
energy supergravity model, supersymmetry is explicitly broken by induced soft
breaking terms of this form but with fewer parameters\cite{sugra}.

We will assume that the spontaneous breaking of the local N=1 supersymmetry is
communicated to the ``visible'' sector by weak gravitational interactions from
some ``hidden'' sector. This spontaneous symmetry breaking of supergravity
manifests itself at low energy as explicit soft breaking terms of global
supersymmetry.

The SUGRA Lagrangian is characterized by two arbitrary functions of the fields,
a real function (the K\"ahler potential) that determines the kinetic terms of
the chiral superfields, and an analytic function, transforming as the symmetric
product of the adjoint representation of the gauge group, that determines the
kinetic terms of the gauge fields. In models with minimal kinetic terms for the
chiral superfields, this leads to a common (gravitino) mass, $m_0$, for all the
scalars of the model.  The presence of non-minimal gauge kinetic terms implies
non-zero masses, $M_l$, for the gauginos at the GUT scale, $M_X$.  By further
assuming gauge coupling unification, we can take the three gaugino masses to be
equal. Furthermore, the trilinear soft couplings $A_u^{ij}$, $A_d^{ij}$, and
$A_e^{ij}$ are all equal to a common value $A_0$. With minimal chiral kinetic
terms, the bilinear soft coupling $B_0$ is related to $A_0$ as $B_0 = A_0 -
m_0$.  This scenario has obvious, desirable features.  First, it is very
predictive since it has a few parameters accounting for thirty-one new masses.
Second, the universal nature of the squark and slepton masses at $M_X$ helps to
avoid the appearance of unwanted flavor changing neutral current (FCNC)
effects.  In fact, one could argue that the absence of FCNCs hints at a
universal mass for the scalars.

All of these couplings will evolve to different values under
the renormalization group.  The complete scalar potential now appears as
\begin{equation}
   V = V_F + V_D + V_{soft} \ ,
\label{vscalar}
\end{equation}
where $V_F$ contains the potential contributions from the $F$-terms
\begin{eqnarray}
   V_F &=& | {\tilde{\overline u}} {\bf Y}_u^{} {\tilde Q}
         + \mu\Phi_d |^2
         + | {\tilde{\overline d}} {\bf Y}_d^{} {\tilde Q}
         + {\tilde{\overline e}} {\bf Y}_e^{} {\tilde L}
         + \mu\Phi_u |^2 \nonumber \\
   & & \mbox{} + | {\bf Y}_u^{} {\tilde Q} \Phi_u |^2
         + | {\bf Y}_d^{} {\tilde Q} \Phi_d
         + {\bf Y}_e^{} {\tilde L} \Phi_d |^2 \nonumber \\
   & & \mbox{} + | {\tilde{\overline u}} {\bf Y}_u^{} \Phi_u
         + {\tilde{\overline d}} {\bf Y}_d^{} \Phi_d |^2
         + | {\tilde{\overline e}} {\bf Y}_e^{} \Phi_d |^2 \ ,
\label{vf}
\end{eqnarray}
and $V_D$ contains the potential contributions from the $D$-terms
\begin{eqnarray}
   V_D &=& {g^{\prime 2}\over2}(\ {1\over6}{\tilde Q}^\dagger{\tilde Q}
           - {2\over3}{\tilde{\overline u}}^\dagger{\tilde{\overline u}}
           + {1\over3}{\tilde{\overline d}}^\dagger{\tilde{\overline d}}
           - {1\over2}{\tilde L}^\dagger{\tilde L}
           + {\tilde{\overline e}}^\dagger{\tilde{\overline e}}
           + {1\over2}\Phi_u^\dagger\Phi_u
           - {1\over2}\Phi_d^\dagger\Phi_d \ )^2 \nonumber \\
   &&\mbox{} + {g_2^2\over8}(\ {\tilde Q}^\dagger{\vec\tau}{\tilde Q}
           + {\tilde L}^\dagger{\vec\tau}{\tilde L}
           + \Phi_u^\dagger{\vec\tau}\Phi_u
           + \Phi_d^\dagger{\vec\tau}\Phi_d \ )^2 \nonumber \\
   &&\mbox{} + {g_3^2\over8}(\ {\tilde Q}^\dagger{\vec\lambda}{\tilde Q}
           - {\tilde{\overline u}}^\dagger{\vec\lambda}^*{\tilde{\overline u}}
           - {\tilde{\overline d}}^\dagger{\vec\lambda}^*{\tilde{\overline d}}
           \ )^2 \ ,
\label{vd}
\end{eqnarray}
where ${\vec\tau}=(\tau_1,\tau_2,\tau_3)$ are the $SU(2)$ Pauli matrices and
${\vec\lambda}=(\lambda_1,\dots,\lambda_8)$ are the Gell-Mann matrices. In
general, one must impose constraints on the parameters to avoid charge and
color breaking minima in the scalar potential.  Some {\it necessary}
constraints have been formulated, such as
%\begin{mathletters}
\label{acontraints}
\begin{eqnarray}
   A_U^2 &<& 3 ( m_Q^2 + m_u^2 + m_{\Phi_u}^2 ) \ ,
   \label{auconstraint}\\
   A_D^2 &<& 3 ( m_Q^2 + m_d^2 + m_{\Phi_d}^2 ) \ ,
   \label{adconstraint}\\
   A_E^2 &<& 3 ( m_L^2 + m_e^2 + m_{\Phi_d}^2 ) \ .
   \label{aeconstraint}
\end{eqnarray}
%\end{mathletters}
\noindent
However, these relations are in general neither {\it sufficient} nor indeed
always {\it necessary} \cite{gunion}.  Their derivation involves very specific
assumptions about the spontaneous symmetry breaking.

\vglue 0.5cm
{\elevenbf \noindent 4. Radiative Electroweak Breaking}
\vglue 0.4cm

An appealing feature of the models we are considering is that they can lead to
the breaking of the electroweak symmetry radiatively
\cite{ibaros,ikkt,agpw,ehnt}. The one loop effective Higgs potential in these
models can be expressed as the sum of the tree level potential plus a
correction coming from the sum of all one loop diagrams with external lines
having zero momenta
\begin{equation}
   V_{1-loop}(\Lambda) =  V_{tree}(\Lambda) + \Delta V_1(\Lambda) \ .
\label{v1loop}
\end{equation}
The one loop correction is given by
\begin{eqnarray}
   \Delta V_1(\Lambda) &=& {1\over64\pi^2} {\rm Str}\{ {\cal M}^4
                  ( \ln{{\cal M}^2\over\Lambda^2} - {3\over2} )\} \nonumber \\
                       &=& {1\over64\pi^2} \sum_p (-1)^{2s_p} ( 2 s_p +  1 )
                  m_p^4 ( \ln{m_p^2\over\Lambda^2} - {3\over2} ) \ ,
\label{dv1}
\end{eqnarray}
where ${\cal M}^2$ is the field dependent squared mass matrix of the model and
$m_p$ is the eigenvalue mass of the $p^{th}$ particle of spin $s_p$. The tree
level part of the potential is
%
%\begin{equation}
%   V_{tree}(\Lambda) = m_1^2 \phi_1^2 + m_2^2 \phi_2^2 - 2 m_3^2 \phi_1 \phi_2
%                   + {1\over8}(g^{\prime 2} + g_2^2)( \phi_1^2 - \phi_2^2)^2
%\label{eq8}
%\end{equation}
%where $\phi_1$ and $\phi_2$ are the neutral components of $\Phi_d$ and
%$\Phi_u$, respectively, and where
%
\begin{eqnarray}
   V_{tree}(\Lambda) &=& m_1^2(\Lambda) \Phi_d^\dagger\Phi_d^{}
                   + m_2^2(\Lambda) \Phi_u^\dagger\Phi_u^{}
                   + m_3^2(\Lambda) (\ \Phi_u\Phi_d + h.c. \ ) \nonumber \\
    &&\mbox{}      + {g^{\prime 2}(\Lambda)\over8}(\ \Phi_u^\dagger\Phi_u^{}
                   - \Phi_d^\dagger\Phi_d^{} \ )^2
                   + {g_2^2(\Lambda)\over8}(\ \Phi_u^\dagger{\vec\tau}\Phi_u^{}
                   + \Phi_d^\dagger{\vec\tau}\Phi_d^{} \ )^2
\label{vtree}
\end{eqnarray}
where
\begin{eqnarray}
   m_1^2(\Lambda) &=& m_{\Phi_d}^2(\Lambda) + \mu^2(\Lambda) \ , \\
   m_2^2(\Lambda) &=& m_{\Phi_u}^2(\Lambda) + \mu^2(\Lambda) \ , \\
   m_3^2(\Lambda) &=& B(\Lambda) \mu(\Lambda) \ .
\label{mvtree}
\end{eqnarray}
The parameters of the potential are taken as running ones, that is, they vary
with scale according to the renormalization group.  The logarithmic term in the
one loop correction is important in making $V_{1-loop}(\Lambda)$ independent of
$\Lambda$ to this order (up to non-field dependent terms).

Given the low energy scale of electroweak breaking, we must use the
renormalization group to evolve the parameters of the potential to a convenient
scale such as $M_Z$ (where the experimental values of the gauge couplings are
usually cited) thereby making this leading log approximation valid.  The exact
scale is not critical as long as it is in the electroweak range.  If we define,
\begin{equation}
   {\overline m}_i^2 = m_i^2 + {\partial\Delta V_1\over\partial v_i^2} \ ,
\label{mbarvtree}
\end{equation}
with $v_1=v_d$, $v_2=v_u$ and
\begin{equation}
   {\partial\Delta V_1\over\partial v_i^2} = {1\over 32\pi^2}\sum_p
   (-1)^{2s_p} ( 2s_p + 1 ) m_p^2 ( \ln{m_p^2\over\Lambda^2} -1 )
   {\partial m_p^2\over\partial v_i^2} \ ,
\label{ddv1dv2}
\end{equation}
then minimization of the potential yields the following two conditions
among its parameters
\begin{equation}
   {1\over2}m_Z^2 = { {\overline m}_1^2 - {\overline m}_2^2 \tan^2\beta \over
                    \tan^2\beta - 1 }  ,
\label{min1}
\end{equation}
where $m_Z^2=(g^{\prime 2}+g_2^2) v^2/2$, $v^2=v_u^2+v_d^2$, and
\begin{equation}
   B \mu = {1\over2}( {\overline m}_1^2 + {\overline m}_2^2 )
                   \sin 2\beta \ ,
\label{min2}
\end{equation}
where $\tan\beta = v_u / v_d$.

Although results based on the tree level potential cannot always be trusted,
one can still use it to get some idea under what conditions electroweak
breaking occurs.  The renormalization group evolution of $m_{\Phi_u}^2$ can be
such that it turns negative at low energies, if the top quark mass is large
enough, whereas $m_{\Phi_d}^2$ runs positive.  From Eq.~(\ref{vtree}), the
scale at which breaking occurs is set by the condition
\begin{equation}
   m_1^2(\Lambda_b) \ m_2^2(\Lambda_b) - m_3^4(\Lambda_b) = 0 \ .
\label{condition1}
\end{equation}
If the free parameters are adjusted properly, then the correct value of the
$Z^0$ mass ($M_Z=91.17$ GeV) can be achieved.

In the tree level analysis, there is another critical scale that must be
considered.  It is evident from Eq.~(\ref{vtree}) that the potential becomes
unbounded from below along the equal field (neutral component) direction, if
\begin{equation}
   m_1^2(\Lambda_s) + m_2^2(\Lambda_s) < 2 m_3^2(\Lambda_s) \ .
\label{condition2}
\end{equation}
Since $m_1^2 m_2^2 - m_3^4 \geq 0$ implies $m_1^2 + m_2^2 \geq 2 m_3^2$,
condition (\ref{condition2}) can only occur at scales lower than condition
(\ref{condition1}), so $\Lambda_s<\Lambda_b$.  From this analysis, one
concludes that the tree level vacuum expectation values (VEVs) of the scalar
fields obtained by minimizing the potential are zero above $\Lambda_b$,
and grow to infinity as one approaches $\Lambda_s$
where the potential becomes unbounded from below.  It follows that the
appropriate scale at which to minimize the tree level potential and evaluate
the VEVs is critical.  This scale must be such that the one loop corrections
of the effective potential may safely be neglected.  Only at such a scale
can the tree level results be trusted.  However, there is more than one scale
involved, and therefore, it is difficult if not impossible to find a scale
at which all logarithms may be neglected.  Indeed, the use of tree level
minimization conditions to compute the VEVs at an arbitrary scale
(e.g., $\Lambda=M_Z$) leads to incorrect conclusions about the regions of
parameter space that yield consistent electroweak breaking scenarios
\cite{gamberini}.  When $\Delta V_1$ is included, however, the value of
$\Lambda$ is not critical as long as it is in the neighborhood of $M_Z$.

Ref.~\cite{gamberini} gives a prescription for arriving at a scale
($\hat\Lambda$) at which
the tree level and the one loop effective potential results for the VEVs
agree.  Three qualitatively different cases are considered.
Following Ref.~\cite{gamberini}, we let $M_{SUSY}$ parametrize the
superparticle thresholds, then the cases can be characterized by the orderings:
(a)~$M_{SUSY}<\Lambda_s<\Lambda_b$,
(b)~$\Lambda_s<M_{SUSY}<\Lambda_b$, and
(c)~$\Lambda_s<\Lambda_b<M_{SUSY}$.  In each case, the prescription is to take
${\hat\Lambda}={\rm max}\{ M_{SUSY},\Lambda_s \}$.  Two cases deserve special
mention.  Case (a) cannot be handled using the tree level analysis because
$v_u,v_d\rightarrow\infty$ near $\Lambda_s$.  Fortunately, phenomenological
bounds rule this case out anyway.  In case (c), there is actually no
electroweak breaking.  For scales below $M_{SUSY}$, the superparticles have
decoupled and the effective theory is not supersymmetric.  Therefore, the
running mass parameters of the potential freeze into their values at
$\Lambda=M_{SUSY}$ at which scale there is no electroweak breaking.  Finally,
it must be emphasized that the apparent violent behavior of the VEVs with
scale in the tree level analysis is an artifact of the approximation.
The only physical potential is the full effective potential, and it either
breaks electroweak symmetry or not.  If it does, then the scalar fields have
non-zero VEVs, and these VEVs are non-zero over all scales varying according
to the anomalous dimension of their respective scalar fields (in the Landau
gauge).

In this paper, we do not rely on the tree level analysis, rather
we incorporate the one loop corrections.  We include the
dominant contributions from the third generation, that is, those of the top
and stop, bottom and sbottom, and tau and stau \cite{ellis,drees}.  The one
loop effective potential is constant against the renormalization group to
this order around the electroweak scale.  We choose the $Z^0$ mass as the
scale at which to evaluate the minimization conditions.  Eqs.~(\ref{min1}) and
(\ref{min2}) can be written
%\begin{mathletters}
\begin{eqnarray}
   \mu^2(M_Z) &=& {{\overline m}_{\Phi_d}^2 -
   {\overline m}_{\Phi_u}^2 \tan^2\beta \over \tan^2\beta - 1 }
   - {1\over2} m_Z^2 \ , \label{min1a} \\
   B(M_Z) &=& { {\overline m}_1^2 + {\overline m}_2^2)\sin2\beta \over
              2 \mu(M_Z) } \ , \label{min2a}
\end{eqnarray}
%\end{mathletters}
where ${\overline m}_{\Phi_{u,d}}^2 = m_{\Phi_{u,d}}^2 +\partial\Delta
V_1/\partial v_{u,d}^2$ and used to solve for $\mu(M_Z)$ and $B(M_Z)$ given
the value of all the relevant parameters at $M_Z$.  We note that the form of
Eq.~(\ref{min1a}) does not fix the sign of $\mu$, and a choice for its
sign must be made ($\mu$ is multiplicatively renormalized).
The right hand sides of these equations implicitly involve the VEV at $M_Z$.
In a consistent scenario it would have the value of $v(M_Z)=174.1$ GeV.
If the parameters are such that $\mu^2<0$, then the scenario is
inconsistent and the electroweak symmetry fails to be broken.

\vglue 0.5cm
{\elevenbf \noindent 5. Boundary Conditions at $M_X$}
\vglue 0.4cm

In this paper, as in the previous one \cite{big} we work in the modified
minimal subtraction scheme (${\overline{\rm MS}}$) of renormalization.  The
parameters of the Lagrangian are not in general equal to any corresponding
physical constant.  For example, in the case of masses, except for those of
the bottom and top quark (see \cite{big}), all other physical masses, $M$,
will be determined from their corresponding running masses by the relation
\begin{equation}
   M = m(\Lambda){\large\vert}_{\Lambda=M} \ .
\label{eq14}
\end{equation}
This equation is easily solved in the course of an integration of the RGEs
for the different masses by noting the scale at which it is valid.
%We have collected the renormalization group $\beta$-functions of the MSSM
%for the gauge and Yukawa couplings to two loops without making any
%approximations in the Yukawa sector in Appendix I.  Also included are the
%one loop $\beta$-finctions for the soft breaking terms.

Because SUGRA models make simplifying predictions about the soft
parameters at the unification scale, we initiate the evolution of the
renormalization group equations at this scale.  It has been demonstrated that
the introduction of supersymmetry leads to gauge coupling unification at
approximately $\sim 10^{16}$ GeV.  Therefore we take $M_X=10^{16}$ GeV, and
evolve down to $1$ GeV, the conventional scale at which the running quark
masses are given \cite{gassleut}.

At the unification scale, $M_X$, all the scalars will have a common mass
%\begin{mathletters}
\begin{eqnarray}
   m_{Q_i}(M_X) &=& m_{u_i}(M_X) = m_{d_i}(M_X) = m_{L_i}(M_X)
   \nonumber \\
                &=& m_{e_i}(M_X) = m_{\Phi_u}(M_X) = m_{\Phi_d}(M_X) = m_0 \ ,
   \label{bcm0}
\end{eqnarray}
%\end{mathletters}
as will the gauginos
\begin{equation}
   M_1(M_X) = M_2(M_X) = M_3(M_X) = m_{1/2} \ .
\label{bcmhalf}
\end{equation}
The prefactors of the trilinear soft scalar terms are equal at $M_X$
\begin{equation}
   A_u^{ij}(M_X) = A_d^{ij}(M_X) = A_e^{ij}(M_X) = A_0 \ .
\label{bca}
\end{equation}
Also we define the bilinear soft scalar coupling and the mixing mass at $M_X$
by $B(M_X) \equiv B_0$, and $\mu(M_X) \equiv \mu_0$.
%\begin{equation}
%   B(M_X) = B_0, \ \mu(M_X) = \mu_0 \ .
%\label{bcbmu}
%\end{equation}

Furthermore, to constrain the parameter space, we will take the bottom and
tau masses equal at $M_X$
\begin{equation}
   m_b(M_X) = m_\tau(M_X) \ .
\label{mbeqmtau}
\end{equation}
This being the best motivated mass relation in supersymmetric grand
unified theories \cite{massrel}.

%--------------------------------------------------------------------------
\begin{table}
\caption{Models}
\begin{center}
\begin{tabular}{lcccc}
\hline
                & $A_0$ & $m_0$ & $m_{1/2}$ & $B_0$ \\
\hline
Strict No-scale & $0$   & $0$   & free       & $0$ \\
No-scale        & $0$   & $0$   & free       & free \\
String Inspired & $0$   & free   & free       & $0$   \\
\hline
\end{tabular}
\end{center}
\end{table}
%--------------------------------------------------------------------------

The present analysis restricts itself to three subclasses of soft symmetry
breaking models.  These have various soft parameters equal to
zero at $M_X$.  The first class of models follows from the
no-scale model \cite{noscale} and has $A_0=m_0=B_0=0$ (strict no-scale model).
In these models, only gaugino masses provide
global supersymmetry breaking.  The second class is the less constraining
no-scale case that has only $A_0=m_0=0$.  A third class we consider with
$A_0=B_0=0$ comes from string derived models.
Table II lists these three possibilities.  In both strict no-scale and string
inspired cases, we must have $B_0=0$.  However, since for us $B_0$ is an
output rather than an input variable, $B_0=0$ results must be inferred from
its behavior upon varying other input parameters.

\vglue 0.5cm
{\elevenbf \noindent 6. Numerical Procedure}
\vglue 0.4cm

In this work as in Ref.~\cite{big}, we will use routines, based on the
``shooting'' method, to solve systems of nonlinear equations.
The method involves making a guess for the solution, then assessing its
merits based on how well the equations are satisfied, given some tolerance.
The process is optimized and iterated until the routine converges on a
solution.

As discussed previously, we start out runs at $M_X$ at which scale
we can make simplifying assumptions about the soft breaking terms based
on various SUGRA models.  This requires that we use the solution routines
to consistently find the $M_X$ values of all known low energy parameters
such as lepton and quark masses and mixing angles and gauge couplings.
This amounts to solving for sixteen unknowns (nine masses, three angles and
a phase, and three gauge couplings).  Alternatively, we could
start our runs at $M_Z$ or $1$ GeV; however, this now requires solving
for sixty-three unknowns (the values of the soft breaking terms at low
energy) that must evolve to just four different values at $M_X$.  The
efficiency of the former method is obvious.

There are seven free parameters in the model we consider.  These are $A_0$,
$B_0$, $m_0$, $m_{1/2}$, $\mu_0$, $\tan\beta$, and $m_t$.  The two
minimization constraints (\ref{min1}) and (\ref{min2}) reduce this
set to five, which are taken to be $A_0$, $m_0$, $m_{1/2}$, $\tan\beta$, and
$m_t$.  In the present framework, $B_0$ and $\mu_0$ will be determined
using the numerical solutions routines in
conjunction with the minimization of the one loop effective potential at
$M_Z$ in the process of evolving from $M_X$ to $1$ GeV.  Minimization at $M_Z$
will give $B(M_Z)$ and $\mu(M_Z)$.  To arrive at $B_0$ and $\mu_0$
(their corresponding values at $M_X$), we employ the solution routine as
follows.  A guess for $B_0$ and $\mu_0$ is made at $M_X$ and then the
parameters of the model are run to $M_Z$ at which scale the evolved value of
$B$ is compared to the minimization output value for $B$ at $M_Z$.  The same
is done for $\mu$.  If the compared values agree to some set accuracy,
then $B_0$ and $\mu_0$ are the required values.  Other analyses that also
extract $B(M_Z)$ and $\mu(M_Z)$ simply evolve these two parameters via
their renormalization group equations back to $M_X$ to find $B_0$ and
$\mu_0$ relying on their near decoupling from the full set of RGEs.
We note that the sign of $\mu$ is not determined from the minimization
procedure, thus we must make a choice for it.  To constrain the parameter
space further, the bottom quark and
tau lepton masses will be taken equal at $M_X$.  This equality is a
characteristic of many SUSY-GUTs.  This constrains the model to four free
parameters, $A_0$, $m_0$, $m_{1/2}$, and $\tan\beta$.  Demanding  that
$m_b(M_X)=m_\tau(M_X)$ and achieving the correct physical
masses for the bottom quark and tau lepton fixes the mass of the top quark
which affects the evolution of the bottom Yukawa significantly.
We shall assume gauge coupling unification, an assumption which appears
reasonable when one considers SUSY models with SUSY breaking scales
$\alt 10$ TeV.

In a complete treatment, the solution routines would be
used to find the precise (similar) values of $\alpha_1$, $\alpha_2$, and
$\alpha_3$ at $M_X$ that
will evolve to the experimentally known values at $M_Z$, however this
increases the CPU time considerably.  We shall therefore sacrifice some
precision in their $M_Z$ values by taking them exactly equal at $M_X$.
This is already a theoretical oversimplification since one does not expect
the gauge couplings to be exactly equal due to threshold effects at the
GUT scale.  We find that for all cases we have studied, the common value
$\alpha_1^{-1}(M_X)=\alpha_2^{-1}(M_X)=\alpha_3^{-1}(M_X)=25.31$ leads to
errors no bigger than $1\%$, $5\%$, and $10\%$ in $\alpha_1(M_Z)$,
$\alpha_2(M_Z)$, and $\alpha_3(M_Z)$, respectively.  This is not so bad
considering that the (combined experimental and theoretical) errors on
$\alpha_3(M_Z)$ from some processes can be as large as $10\%$ \cite{big}.

It is well known that there is a fine tuning problem inherent in the
radiatively induced electroweak models.  For certain values of the
parameters, the top quark mass must be tuned to an ``unnaturally'' high
degree of accuracy to achieve the correct value of $M_Z$.  This problem is
generally handled by rejecting models that require ``too much'' tuning.  The
amount of tuning is usually defined quite arbitrarily.  The usual procedure
is to define fine tuning parameters
\begin{equation}
   c_i = | {x_i^2\over M_Z^2} {\partial M_Z^2\over\partial x_i^2} | \ ,
\label{ftparam}
\end{equation}
where $x_i$ are parameters of the theory such as $m_0$, $m_{1/2}$,
$\mu$, or $m_t$.  One then demands that the $c_i$ be less than some
chosen value that is typically taken to be $10$.

We have analyzed to
some extent the differences in using the tree level {\it vs.} one loop
effective potential.  The basis for the ``theoretical'' fine tuning problem
can be seen, if one makes some simplifying assumptions, in the dependence
of $M_W$ on the top quark Yukawa coupling, $y_t$, \cite{nano}
\begin{equation}
   M_W \sim M_X e^{-1/y_t^2} \ .
\label{finetuning}
\end{equation}
We remark that this fine tuning problem is exacerbated if one uses only the
tree level analysis of the potential.  The vacuum expectation value coming
from the minimization conditions
of the tree level potential changes rapidly from $0$ to infinity over the
interval $(\Lambda_s,\Lambda_b)$.  Using the prescription of
Ref.\cite{gamberini} for the scale ${\hat \Lambda}$ at which to adequately
minimize the tree level potential, to extract $v({\hat\Lambda})$,
and thereby to arrive at a value for $M_Z$, one finds that although a small
variation in $y_t(M_X)$ may lead to a small variation in $\Lambda_b$, the
steepness in the tree level VEV can lead to a large variation in the value of
$v({\hat\Lambda})$ and therefore in $M_Z$.  Hence, in the tree level analysis,
solutions which may be within the bounds of the ``theoretical'' fine tuning
may nevertheless display a fine tuning aspect because of this ``tree level''
fine tuning of $\hat\Lambda$.
However, our use of the one loop effective potential
(i.e., including $\Delta V_1$) stabilizes the VEV around the $M_Z$ scale
and this particular fine tuning goes away.  The true VEVs depend on scale
through wave function renormalization effects which are never large as can be
seen from the form of the renormalization group equations for the VEVs.

In this analysis, we shall also reject solutions based on fine tuning
considerations; however, our method differs somewhat from the usual one in
that it is incorporated in the solution routine described above.  The routine
is an iterative one which determines the convergence properties of the
solution.  Very slow convergence reflects an inherent fine tuning.  Therefore,
if the convergence is too slow, we will reject the solution.  Effectively we
are rejecting any solution which the computer cannot pinpoint within an
allotted number of iterations.

Given values for $A_0$, $m_0$, $m_{1/2}$,
$\tan\beta$, and ${\rm sign}(\mu)$, the solution routines search for the
values of $v(M_X)$, $m_{b,\tau}(M_X)$, $m_t(M_X)$, $B_0$, and $\mu_0$.
The process by which $B_0$ and $\mu_0$ are found was described above.  The
remaining three parameters are determined similarly.  The routine makes a
guess for $v(M_X)$, $m_{b,\tau}(M_X)$, and $m_t(M_X)$, then the full
renormalization group equations are evolved to $1$ GeV calculating
superparticle threshold masses in the process and minimizing the one loop
effective potential at $M_Z$.  The merits of the guess for $v(M_X)$,
$m_{b,\tau}(M_X)$, and $m_t(M_X)$ is assessed by comparing the resulting
values of $M_Z$, $m_\tau(1~{\rm GeV})$, and $m_b(1~{\rm GeV})$ with the
expected ones.  The process is iterated until the correct values are
achieved to within some tolerance.

\vglue 0.5cm
{\elevenbf \noindent 7. Thresholds}
\vglue 0.4cm

In the minimal low energy supergravity model
being considered, the super particle spectrum is no longer degenerate as
in the simple global supersymmetry model in which all the super particles
are given a common mass, $M_{SUSY}$.  In the simple case, one makes one
course correction in the renormalization group evolution at $M_{SUSY}$.
In the model with soft symmetry breaking, the nondegenerate spectrum should
lead to various course corrections at the super particle mass thresholds.
To this end, the renormalization group $\beta$ functions must be cast in a
new form which makes the implementation of the thresholds effects (albeit
naive) evident.  Since the ${\overline {\rm MS}}$
renormalization group
equations are mass independent, particle thresholds must be handled using
the decoupling theorem \cite{appel}, and each super particle mass
has associated with it a boundary between two effective theories.  Above
a particular mass threshold the associated particle is present in the
effective theory, below the threshold the particle is absent.

The simplest way to incorporate this is to (naively) treat the thresholds as
steps in the particle content of the renormalization group $\beta$ functions.
This method is not always entirely adequate.  For example, in the case of the
$SU(2)$ gauge coupling there will be scales in the integration process at which
there are effectively a half integer number of doublets using this method.
We believe, nevertheless, that this method does yield the correct, general
behavior of the evolution.  It is a simple means of implementing the smearing
effects of the non-degenerate super particle spectrum.  The determination of
the spectrum of masses is done without iteration as is common in other
analyses.  Our method deduces the physical masses by solving the equation
$m(\Lambda)=\Lambda$ for each superparticle in the process of evolving from
$M_X$ to
$1$ GeV.  The usual iterative method requires several runs to find a consistent
solution.

\vglue 0.5cm
{\elevenbf \noindent 8. Analysis}
\vglue 0.4cm

The tremendous computing task involved in analyzing the full parameter space
of the soft symmetry breaking models, using the methods described as
designed, would be far too time consuming given the computing facilities
available to us.  Therefore, in the following analysis, some simplifications
will be made in the procedural method.  First, only the heaviest family of
quarks and leptons will have non-zero mass.  This will cut down on the
CPU time required for the solution routines to consistently find their
values at $M_X$.  Second, as stated previously, the
value of the strong coupling at $M_Z$ will be allowed to vary from its
central value of $.113$ by at most $10\%$.  This translates into a similar
error in the bottom quark mass.  Third, the allotted number of Runge-Kutta
steps, involved in numerically integrating the renormalization group
equations, will be cut down to $\sim 100$.

Our method involves four input parameters $A_0$, $m_0$, $m_{1/2}$, $\tan\beta$
(and the ${\rm sign}(\mu)$).  The output is $B_0$, $\mu_0$, $M_t$, $M_h$,
and all the masses of the extra particles associated with the MSSM.
Efficient use of CPU time required that we proceed as follows.  For a given
model, our initial exploration of the parameter space was performed in a
coarse grained fashion.  $\tan\beta$ was most commonly coarse grained as $2$,
$5$, and $10$, with only some runs involving higher values (e.g., $15$, $20$).
The other three input parameters were varied in steps of $50$ and $100$ GeV.
Values larger than $\sim 500$ GeV were rarely ever used.  We subsequently
narrowed down on the allowed hyperenvelope by fine graining around the edges
of the expected region (based on the coarse graining results).

Our raw data consists of those runs which satisfy the following two criteria.
The first is consistent electroweak breaking; that is, the correct value of
$M_Z$ is achieved from the minimization of the one loop effective potential
with $\mu^2(M_Z)>0$.  The second criterion is no fine tuning, as implemented in
our method (see Section 6).  The solution routines employed return a numbered
code representing the convergence properties of the solution which we use to
screen the runs.

The raw data is then progressively filtered based on three physical
constraints.  R-parity is a discrete symmetry which distinguishes particles
from superparticles by assigning a $+1$ to particles and a $-1$ to the
superparticles. Conserved R-parity requires the existence of a stable lightest
supersymmetric particle (LSP).  Astrophysical considerations indicate that the
LSP must be neutral and colorless.  Cosmological considerations based on the
LSPs contribution to the density of the universe indicate that it must have a
mass less than $\sim 200$ GeV \cite{arnnat,robros,lnz}.
First, points in parameter space that lead to
LSPs other than neutralinos with masses less than $\sim 200$ GeV are cut.
Second, flavor changing neutral current bounds are used to reject runs in
which intergenerational splitting of squark and slepton masses is too large.
Third, experimental limits on the masses of the superparticles are used as
another criterion to reject runs.

\vglue 0.5cm
{\elevenbf \noindent 9. Results}
\vglue 0.4cm

In the following, we discuss the results of our analysis of the three classes
of models listed in Table II.  Because we have performed a coarse grained
study, our results should only be
considered qualitatively valid.  Based on these, we hope to be able to
ascertain the general trends in the data, and make some general predictions
about the feasibility of the models considered.

Because the GUT inspired constraint Eq.~(\ref{mbeqmtau}) is enforced
in this analysis, the results will depend on the mass of the bottom quark.
Namely, lower bottom quark masses require larger values of the top quark mass
to satisfy this relation.
Most results will be reported for the case $m_b(1~{\rm GeV})=6.00$ GeV,
but lower mass ($5.70$ GeV) and higher mass ($6.33$ GeV) cases were also
studied.  The running value of $6.00$ GeV for $m_b(1~{\rm GeV})$ corresponds
to a physical bottom mass $M_b=4.85\pm .15$ GeV, with the uncertainty coming
from the error in the strong coupling, as discussed above.
%\newpage
\vglue 0.2cm
{\elevenit\noindent 9.1. No-scale Case}
\vglue 0.1cm

In all cases considered, the mass of the LSP, when it is a neutralino, is
observed to be correlated with the value of $m_{1/2}$.  Therefore, we find
that $m_{1/2}$ cannot be taken too large ($\alt 400$ GeV).  In the no-scale
case, we present plots of $M_t$ vs. $m_{1/2}$ for three values of
$m_b(1~{\rm GeV})$ and containing all points satisfying the various
criteria outlined in Section 8.  Figure 1 is such a plot for
$m_b(1~{\rm GeV})=6.00$ GeV.  The ``right edge'' of the envelope is defined
by points whose neutralino LSPs are just slightly heavier than the lightest
charged superparticle (usually a ${\tilde\tau}_{\scriptscriptstyle R}$ for
us).  The ``left edge'' defines the threshold of consistent electroweak
breaking.  The top and bottom edges are set by the requirement that
Eq.~(\ref{mbeqmtau}) hold.  The range of $\tan\beta$ considered leads to
definite lower and upper bounds on $M_t$ \cite{summer91,kln,anan,giveon}.

%%%%%%%%%%%%%%%%%figure
%\vskip 5truein
%{\tenrm\baselineskip=12pt
%Figure 1. Plot of $M_t$ vs. $m_{1/2}$ for $\tan\beta=2,5,10,15$ and
%$m_b(1~{\rm GeV})=6.00$ GeV.}
%\vglue 0.4cm
%\newpage
%\noindent

In Figs.~2 and 3, we display similar plots with lower ($m_b(1~{\rm GeV})=5.70$
GeV) and higher ($m_b(1~{\rm GeV})=6.33$ GeV) bottom quark masses.  The
previously noted dependence of the data on $M_b$ is evident from these
figures.
%%%%%%%%%%%%%%%%%figure
%\vskip 6truein
%{\tenrm\baselineskip=12pt
%Figure 2. Same as Fig.~1 but with $m_b(1~{\rm GeV})=5.70$ GeV.}
%\vglue 0.4cm
%\newpage
In Fig.~3, we change the value of the bottom quark mass to illustrate the
dependence of our results on its value.
%%%%%%%%%%%%%%%%%figure
%\vskip 6truein
%{\tenrm\baselineskip=12pt
%Figure 3. Same as Fig.~1 but with $m_b(1~{\rm GeV})=6.33$ GeV.}
%\vglue 0.4cm

{}From these, we note that the available range of $m_{1/2}$ decreases with
increasing $M_b$.
Fig.~1 indicates that $190~{\rm GeV}\alt m_{1/2}\alt 265~{\rm GeV}$.
We can draw no conclusions, however, about the value of $\tan\beta$.
We can conclude from the $m_b(1~{\rm GeV})=6.00$ GeV case that
$85 \alt M_t \alt 132$ GeV.  From similar plots involving $M_h$, we conclude
that $25 \alt M_h \alt 78$ GeV.
%\newpage
In Fig.~4, we display the familiar dependence
of $M_t$ on $\tan\beta$ for a particular value of $m_{1/2}$ in the allowed
range.

%%%%%%%%%%%%%%%%%figure
%\vskip 6truein
%%{\tenrm\baselineskip=12pt
%Figure 4. Plot of $M_t$ vs. $\tan\beta$ in the no-scale case with
%$m_{1/2}=240$ GeV and $m_b(1~{\rm GeV})=6.00$ GeV.}
%\vglue 0.4cm

%\noindent
The official experimental lower bound on the top quark mass is $108$ GeV.
The figures indicate that the top quark cannot have a mass greater than
$\sim 132$ GeV in this model, if $m_b(1~{\rm GeV})=6.00$ GeV.  This upper bound
is raised to $\sim 160$ GeV if $m_b(1~{\rm GeV})=5.70$ GeV, and the model is
ruled out, if $m_b(1~{\rm GeV})=6.33$ GeV.
%\newpage
Figure 5 is similar to Fig.~1, but we have chosen sign of $\mu$ negative.
The allowed region is displaced down with respect to the positive $\mu$
case with the upper bound on $M_t$ now $113$ GeV and very close to the
experimental limit.
%%%%%%%%%%%%%%%%%figure
%\vskip 6truein
%{\tenrm\baselineskip=12pt
%Figure 5. Same as Fig.~1 but with ${\rm sign}(\mu)=-$.}
%\vglue 0.4cm
%\newpage

\vglue 0.2cm
{\elevenit\noindent 9.2. Strict No-scale Case}
\vglue 0.1cm

The results of the strict no-scale case must be interpolated from the
no-scale results, because $B_0$ is not an input parameter in our procedural
method.  Therefore, we plot in Fig.~6 $M_t$ vs. $B_0$ and deduce the $M_t$
bounds from slicing the data along $B_0=0$.  Admittedly, the relatively
small number of points makes the perimeter of the region unclear in some
areas.  We get approximately $90 \alt M_t \alt 127$ GeV.  For $M_h$, we
find $40 \alt M_h \alt 78$ GeV with an uncertainty in the lower bound of
$\sim 10$ GeV due to lack of definition in the lower end of the envelope.
Inspection of the data indicates that $3.5 \alt \tan\beta \alt 9$.  Thus,
it appears that $\tan\beta$ cannot be too small or too large to accommodate
the strict no-scale case.
%%%%%%%%%%%%%%%%%figure
%\vskip 5truein
%{\tenrm\baselineskip=12pt
%Figure 6. Plot of $M_t$ vs. $B_0$ for points with $A_0=m_0=0$ used to
%interpolate to $B_0=0$ and deduce the $M_t$ range in this case.}
%\vglue 0.4cm
%\newpage
%\noindent

Finally, in Table 3, we display the spectrum of superparticle masses for a
representative strict no-scale scenario with $\tan\beta=8.3$ and $m_{1/2}=240$
GeV.  The masses are in GeVs. There are several interesting features. For
instance, the LSP is the photino in this case and has a mass of $92$ GeV.  It
is too light to account for the dark matter. The top quark mass is just above
the experimental limit at $126$ GeV, and the Higgs boson mass is $77$ GeV,
which underlines one of the signatures of these scenarios: the Higgs is
typically light, so that it will be found in two photon modes at SSC. At any
rate, it illustrates the predictive power of this possibility.

%--------------------------------------------------------------------------
\begin{table}
\caption{Strict No-scale Scenario}
\begin{center}
\begin{tabular}{lc}
\hline
$A_0$             & $0$   \\
$m_0$             & $0$   \\
$m_{1/2}$         & $240$ \\
$\mu_0$           & $171$ \\
$\tan\beta$       & $8.3$ \\
${\rm sign}(\mu)$ & $+$   \\
\hline
$M_t$             & $126$ \\
\hline
${\tilde d}$      & $473-506$ \\
${\tilde u}$      & $390-532$ \\
\hline
${\tilde e}$      & $94-176$  \\
${\tilde\nu}_L$   & $180$     \\
\hline
${\tilde\gamma}$           & $92$         \\
${\tilde Z}$, ${\tilde W}$ & $153$, $147$ \\
${\tilde g}$               & $557$        \\
\hline
${\tilde H}^0$   & $213$, $271$ \\
${\tilde H}^\pm$ & $270$        \\
\hline
$h$, $H$     & $77$, $158$  \\
$H^\pm$, $A$ & $235$, $222$ \\
\hline

\end{tabular}
\end{center}
\end{table}
%--------------------------------------------------------------------------
\vglue 0.2cm
{\elevenit\noindent 9.3. String Inspired Case}
\vglue 0.1cm

As in the strict no-scale, our results for the string inspired case
necessitates interpolating $A_0=0$ data to $B_0=0$.  Once again we find
the perimeter of the allowed region is not well defined everywhere.  Hence,
our results are only qualitative.  Figure 7 is similar to Fig.~6, but in
this case we only fix $A_0=0$.
%\newpage
The strict no-scale case is a special case of the string inspired one,
therefore the $127$ GeV top quark mass upper bound is not expected to
decrease but rather to increase in this case.
Slicing along $B_0=0$ yields
$85 \alt M_t \alt 140$ GeV.  Similarly, for the light Higgs we get
$40 \alt M_h \alt 80$ GeV.  The data indicates in this case, as in the
strict no-scale case, that there is a lower bound on $\tan\beta$ of $\sim 3$.
%%%%%%%%%%%%%%%%%figure
%\vskip 5truein
%{\tenrm\baselineskip=12pt
%Figure 7. Same as Fig.~6 but for points with $A_0=0$ only.}
%\vglue 0.4cm

\vglue 0.5cm
{\elevenbf \noindent 10. Conclusions}
\vglue 0.4cm

Minimal low energy supergravity models were considered.  It is quite remarkable
that for as simple a breaking of supersymmetry as that offered by the strict
no-scale model, we can reproduce the standard model results, including the
appealing feature of automatic breaking of the electro-weak symmetry. The
importance of the value of the top quark mass to these schemes cannot be
understated, since we found in our study of specific models, rather restrictive
upper bounds for the top quark mass.  No-scale models in which only gaugino
masses provide global supersymmetry breaking yield top quarks with masses less
than $\sim 127$ GeV.  The results are sensitive to the value of the bottom
quark mass.  Lower bottom quark masses, within the experimental uncertainty,
lead to higher top quark upper bounds.  In these models, the ratio of vacuum
expectation values of the two Higgs fields is expected to be larger than $\sim
70^\circ$.

Although the perimeter of the allowed regions were often fuzzy, we
could, nevertheless, draw some general conclusions from our results.
For all our runs, with no restrictions on the soft terms, we find for
the top quark $M_t \alt 186$ GeV and for the light Higgs boson
$M_h \alt 100$ GeV .

\vglue 0.5cm
{\elevenbf \noindent 11. References}
\vglue 0.4cm

\vglue 0.2cm

\vglue 0.5cm
{\elevenbf \noindent 12. Figure Captions}
\vglue 0.4cm

{\elevenrm\baselineskip=12pt
Figure 1. Plot of $M_t$ vs. $m_{1/2}$ for $\tan\beta=2,5,10,15$ and
$m_b(1~{\rm GeV})=6.00$ GeV.}
\vglue 0.4cm
{\elevenrm\baselineskip=12pt
Figure 2. Same as Fig.~1 but with $m_b(1~{\rm GeV})=5.70$ GeV.}
\vglue 0.4cm
{\elevenrm\baselineskip=12pt
Figure 3. Same as Fig.~1 but with $m_b(1~{\rm GeV})=6.33$ GeV.}
\vglue 0.4cm
{\elevenrm\baselineskip=12pt
Figure 4. Plot of $M_t$ vs. $\tan\beta$ in the no-scale case with
$m_{1/2}=240$ GeV and $m_b(1~{\rm GeV})=6.00$ GeV.}
\vglue 0.4cm
{\elevenrm\baselineskip=12pt
Figure 5. Same as Fig.~1 but with ${\rm sign}(\mu)=-$.}
\vglue 0.4cm
{\elevenrm\baselineskip=12pt
Figure 6. Plot of $M_t$ vs. $B_0$ for points with $A_0=m_0=0$ used to
interpolate to $B_0=0$ and deduce the $M_t$ range in this case.}
\vglue 0.4cm
{\elevenrm\baselineskip=12pt
Figure 7. Same as Fig.~6 but for points with $A_0=0$ only.}

\end{document}